\documentclass[aps,prl,twocolumn]{revtex4-1}
\usepackage{amsmath,amssymb}
\usepackage{graphicx}
\usepackage[dvipsnames]{xcolor}
\usepackage{comment}
\usepackage[overload]{empheq}
\usepackage{upgreek}
\usepackage{lmodern,pdftexcmds}
\usepackage[sort&compress]{natbib}
\usepackage{cases}

\newcommand{\aleq}{\mbox{\
\raisebox{-.9ex}{$\stackrel{\textstyle<}{\sim}$}\ }}
\newcommand{\ageq}{\mbox{\
\raisebox{-.9ex}{$\stackrel{\textstyle >}{\sim}$}\ }}

\begin{document}

\title{Exceptionally Dense and Resilient Polydisperse Disk Packings}

\author{Sangwoo Kim}
\affiliation{Institute of Mechanical Engineering, École Polytechnique Fédérale de Lausanne (EPFL), CH-1015 Lausanne, Switzerland}

\author{Sascha Hilgenfeldt}
\affiliation{Mechanical Science and Engineering, University of Illinois, Urbana-Champaign, Illinois 61801, USA}

\date{\today}

\begin{abstract}
Understanding the way disordered particle packings transition between jammed (rigid) and unjammed (fluid) states is of both great practical importance and strong fundamental interest. The values of critical packing fraction (and other state variables) at the jamming transition are protocol dependent. Here, we demonstrate that this variability can be systematically traced to structural measures of packing, as well as to energy measures inside the jamming regime. A novel
generalized simultaneous particle swap algorithm constructs overjammed states of desired energy, which upon decompression lead to predictable critical packing fractions.
Thus, for a given set of particle sizes, states with extraordinarily high critical packing fractions can be found efficiently, which sustain substantial shear strain and preserve their special structure over the entire jammed domain. The close relation revealed here between the energy landscape of overjammed soft-particle packings and the behavior near the jamming transition points towards new ways of understanding and constructing disordered materials with exceptional properties.
\end{abstract}

\maketitle

Particle based models are used to study a wide range of  materials, including glasses, colloidal aggregates, foams, emulsions, and granular matter. The packing fraction $\phi$ of particles  is an important indicator of mechanical properties of the packing, including the rigidity transition known as jamming \cite{OHern03,Hecke09,Behringer18}. At the jamming transition, constituent particles form a percolating contact network, satisfying isostatic conditions \cite{Ellenbroek15}. The material behavior near the transition point has been described in the language of  critical phenomena. \cite{OHern03,Ellenbroek06,Liu10,Olsson07,Sartor21} 

\begin{figure}[!t]
\centering
\includegraphics[width=.45\textwidth]{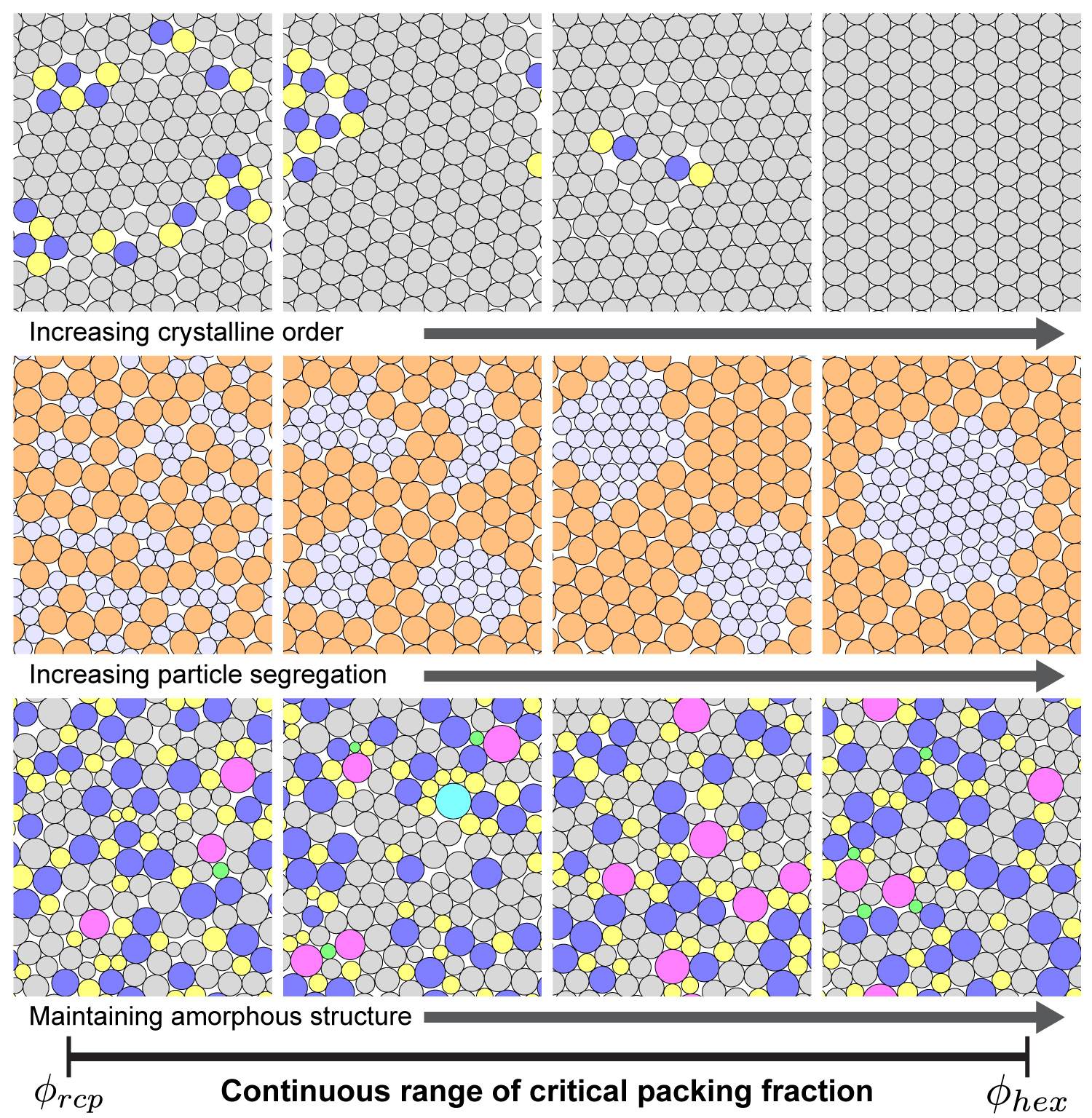}
\hfill
\caption{Schematic of critically jammed structures for monodisperse (top), bidisperse (center), and polydisperse (bottom) frictionless disks. Monodisperse and polydisperse disks are colored by number of neighbors,vbidisperse disks by size.}
\label{figmonobipoly}
\end{figure}

Recent research has shown, however, that the critical packing fraction at the jamming transition $\phi_c$ is not unique and depends on the preparation protocol of the packing, even for simple systems like frictionless hard circular disks in two dimensions or frictionless hard spheres in three dimensions \cite{Chaudhuri10,hopkins2013disordered,schreck2011tuning,Ozawa17,brouwers2023geometric}. To understand mechanical properties of the material, it is crucial to understand and predict $\phi_c$. Figure~\ref{figmonobipoly} illustrates structural changes associated with different $\phi_c$. The packing fraction of the hexagonal crystal $\phi_{hex}=0.9069\dots$ is proved to be the highest $\phi_c$ for monodisperse disks and also provides an upper bound on $\phi_c$ of bidisperse systems with sizes more similar than a threshold value \cite{toth1972covering,heppes2003some}, which structurally corresponds to size segregation. In both mono- and bidisperse systems, lower $\phi_c$ indicates stronger positional disorder (increased defect density and smaller-scale segregation, respectively). Bidisperse systems with larger size contrast can avoid crystallization entirely and in many simulation protocols settle on a critical density $\phi_{rcp}\approx 0.84$ of "random close packing", although lower-density random packings are possible with specialized protocols in monodisperse systems \cite{Atkinson14}. 

Systems with significant continuous size polydispersity are less often studied, but avoid long-range order very effectively. Again the vast majority of packing protocols yield $\phi_c\approx \phi_{rcp}$. We show here that, for a wide range of given particle size distributions, such fully polydisperse packings can be found with $\phi_c$ close to $\phi_{hex}$, without any long-range order (Fig.~\ref{figmonobipoly}). 
We present specialized protocols finding these unusual states efficiently and demonstrate that they indeed have extraordinary mechanical properties and structural robustness.

In contrast to many common protocols of the Lubachevsky-Stillinger (LS) type \cite{lubachevsky1990geometric}, which approach the critical jamming point from unjammed (fluid) states by incrementally increasing packing fraction, we instead use 
polydisperse soft particles, generating overjammed states ($\phi>\phi_c$), which we relax by incrementally decreasing $\phi$ following \cite{Xu05,Desmond09}. Crucially, for these systems a simple, computationally inexpensive protocol allows us to select special overjammed states that at transition generate jamming states of exceptionally high $\phi_c$.

\begin{figure}[tp!]
\centering
\includegraphics[width=.47\textwidth]{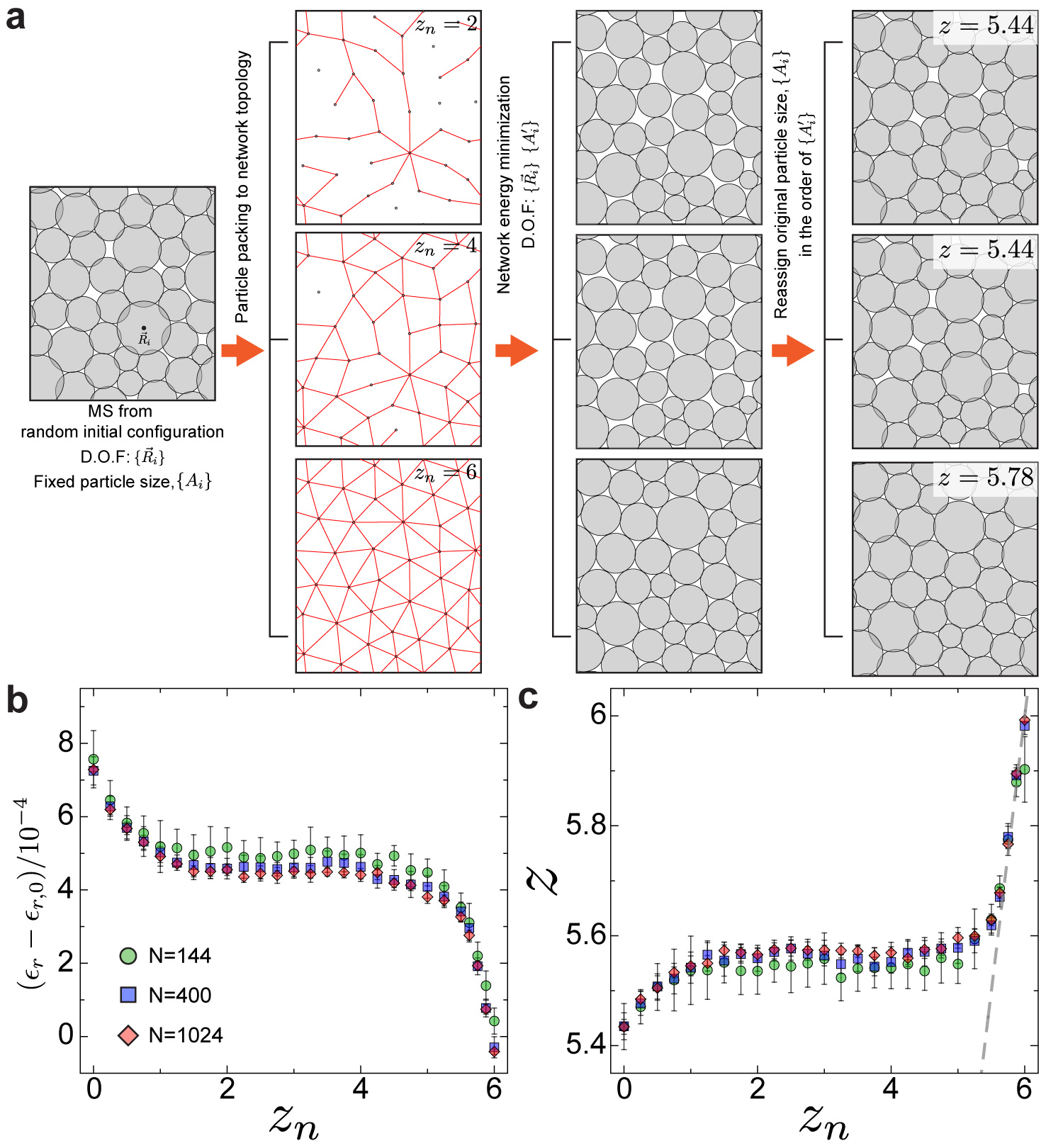}
\hfill
\caption{\textbf{a} Schematic of the generalized angle swap simulation protocol to generate the entire range of metastable state energies for overjammed packings. \textbf{b} Correlation between the average network coordination number $z_n$ and the corresponding metastable state energy $\epsilon_r$ at $\phi=1$, \textbf{c} correlation between $z_n$ and mean contact number of the overjammed particle packing $z$. }
\label{figzn}
\end{figure}

In overjammed soft-particle packings  (Fig.\ref{figzn}a),  particles in contact interact through pair potentials and the entire system is described (at zero temperature) by a multidimensional energy landscape whose degrees of freedom are the particle positions.
An initial configuration relaxes to one of a huge number of mechanical equilibrium states, representing local minima of the energy landscape (metastable states, MS), 
which correspond to inherent structures in the parlance of the glass literature \cite{Stillinger84,Debenedetti01}. In recent work, the authors showed that deeply overjammed MS ($\phi=1$) with distinct energy levels exhibit distinct structural features as well as mechanical behavior, and that structural measures of the states can be utilized to efficiently construct low energy MS \cite{Kim22}.
Here, we generalize the efficient construction of such states to all MS energies and all jammed packing fractions, reaching very large $\phi_c$ without having to assume specialized particle size distributions.

We model the interaction between particles $i$ and $j$ by a repulsive harmonic potential $V(r_{ij})$ (cf.~\cite{OHern03}) with normalized spring constant, namely
\begin{equation}
    V(r_{ij})=\frac{1}{2}\left(1-\frac{r_{ij}}{\sigma_{ij}}\right)^2 H\left(1-\frac{r_{ij}}{\sigma_{ij}}\right)\,.
    \label{eq:vr}
\end{equation}
Here, $r_{ij}$ is the distance between centers of particles $i$ and $j$, $\sigma_{ij}=\sigma_i+\sigma_j$ is an equilibrium distance equal to the sum of particle radii, and $H(\cdot)$ is the Heaviside step function. For circular particles, radii $\sigma_i=\sqrt{A_i/\pi}$ follow from areas  $A_i$, whose distribution we fix.
 We stress that the results presented are independent of the particular functional form (\ref{eq:vr}).

The total non-dimensional energy $\epsilon_r$ of a given metastable state and the reference energy $\epsilon_{r,0}$ of a monodisperse hexagonal packing at packing fraction $\phi$ are then
\begin{equation}
\epsilon_r=\frac{1}{3N}\sum_{i<j}V(r_{ij})\,,
\end{equation}
\begin{equation}
    \epsilon_{r,0}=\frac{1}{2}\left(1-\sqrt{\frac{\pi}{2\sqrt{3}\phi}}\right)^2\,.
\end{equation}

In \cite{Kim22}, the authors introduced the angle swap algorithm for strongly overjammed packings of $\phi=1$ (far from the jamming transition), in which a single simultaneous particle size swap is sufficient to anneal high energy metastable states to low energy states. While this is a very efficient algorithm for finding very low energy configurations, it cannot construct metastable states at intermediate energy. To overcome this limitation, we generalize the angle swap algorithm using the concept of particle-based networks (Fig.~\ref{figzn}a).
 An initial configuration is constructed at $\phi=1$ by generating $N$ particle sizes $\{A_i\}$ randomly from a given size distribution and placing the particles at random positions $\{\vec{R}_{i}\}$ within a periodic box of fixed size. We use Gamma, log-normal, and uniform area distributions and quantify their width by the coefficient of variation $c_A$. 
 We restrict the width between $c_A=0.2$ and  $c_A=0.5$ to avoid hexagonal ordering (crystallization) for low polydispersity and Apollonian fractal-like packings for high polydispersity. In the following, we will focus on Gamma-distributed areas of $c_A=0.4$, but the results hold for all cases mentioned (see Supplementary Information for details).
 
The initial configuration is annealed to the nearest metastable state using the FIRE algorithm \cite{Bitzek06}. The resulting MS overjammed packing is then translated to a network system of defined topology by identifying all neighboring pairs by radical tessellation \cite{Okabe92} along with their normalized distances, $r_{ij}/\sigma_{ij}$. 
Identifying particle positions with nodes of the network, network bonds are added sequentially in order of increasing normalized neighbor distance until the average coordination number of the network is larger than or equal to a target value $z_n$. To find equilibria of the network, we associate a harmonic potential energy with the bonds, but there are two major differences to the particle system: (1) the network topology is fixed while the contact topology of particle packings can change according to the change of particle positions; (2) node particle sizes, $\{A_i\}$, are considered as degrees of freedoms in addition to node positions. This latter treatment is reminiscent of recent work simultaneously optimizing particle positions and sizes for particle packings \cite{Hagh21}, though that work does not enforce fixed network topology. We anneal the network system to the nearest metastable state, resulting in a new set of particle sizes $\{A^{\prime}_i\}$. The initial particle sizes $\{A_i\}$ are then reassigned to particle positions in the order of the $\{A^{\prime}_i\}$ sizes, effecting a simultaneous size swap for the entire particle system. This approach extends the conventional swap algorithm, widely used to expedite the search for equilibrium states in glass simulations \cite{Grigera01, Ninarello17, Brito18}.
Finally, this new packing configuration of the original set of particles is annealed to the nearest metastable state. The entire procedure involves only three energy minimization steps, which take up most of computation time.  

\begin{figure}[!t]
\centering
\includegraphics[width=.47\textwidth]{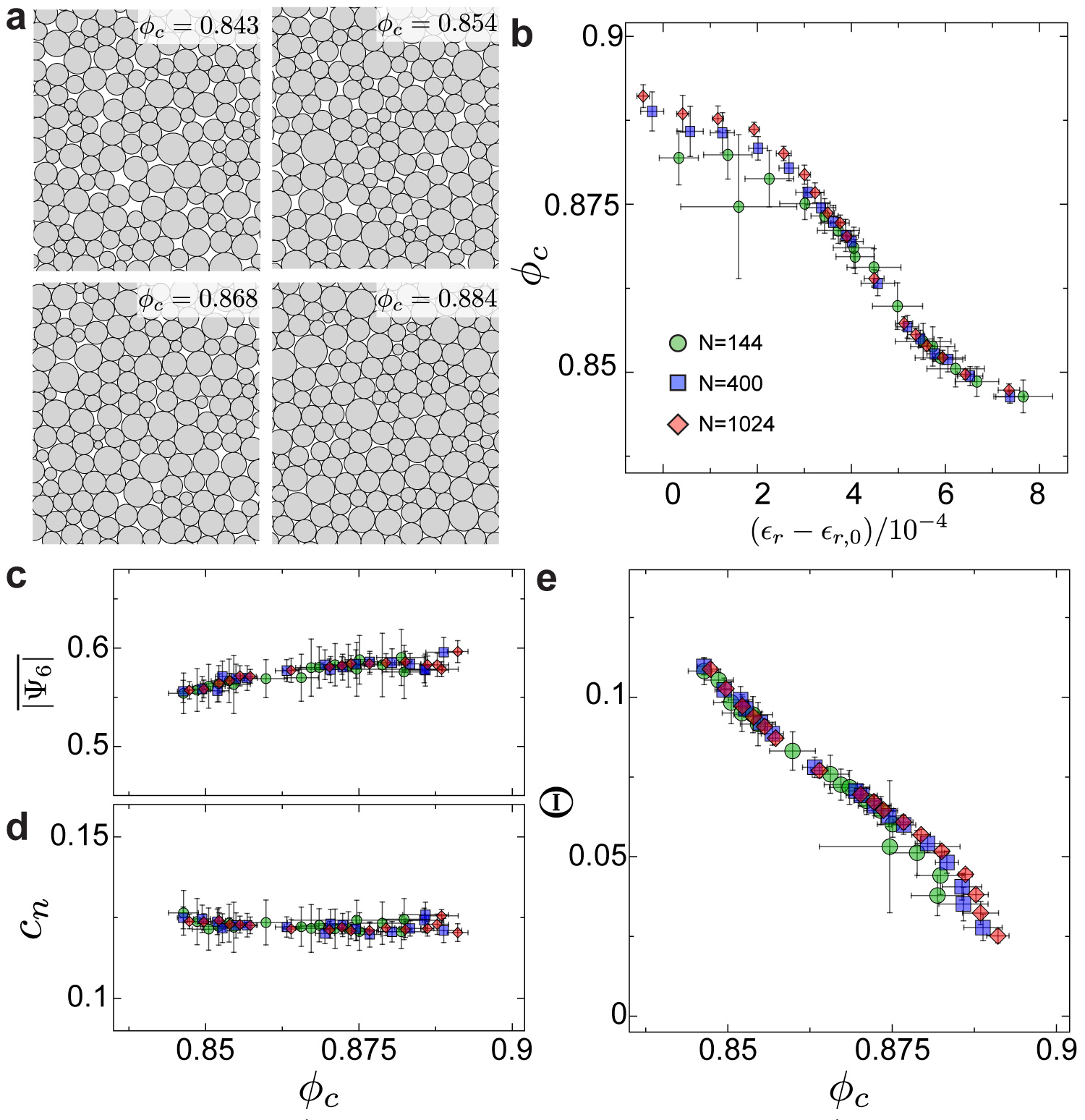}
\hfill
\caption{\textbf{a} Representative jamming configurations for different critical packing fractions. \textbf{b} The packing fraction $\phi_c$ of critically jammed states is strongly predicted by the energy level of corresponding overjammed packings. Symbols represent different sizes of individual simulation systems. \textbf{c} Hexagonal orientation order parameter $
\left<\Psi_6\right>$ and \textbf{d} coefficient of variation of radical tessellation topology $c_n$ do not predict $\phi_c$ well, while \textbf{e} the angle order parameter $\Theta$ is an excellent predictor. }
\label{fig3}
\end{figure} 

This algorithm efficiently generates metastable states of overjammed packings for the entire energy range, with $z_n$ as a control parameter (Fig.~\ref{figzn}b). The energy is insensitive to $z_n$ in the range $1\aleq z_n\aleq5$, but sharply decreases from the highest values at $z_n=0$ and again towards the lowest values at  $z_n=6$. The approach remains effective for a range of overjammed $\phi$ values (see Supplementary Information).
The mean number of particle contacts $z$ in the final metastable state is similarly controlled by $z_n$ (Fig.~\ref{figzn}c), in agreement with the reported negative correlation between $z$ and MS energy \cite{Kim22}. Figure~\ref{figzn}b,c also demonstrates that the correlation is independent of the system size. 
The mean displacement of a particle between an initial MS and a final MS is far less than an average particle size in most cases, implying that local adjustment of particle positions and sizes is sufficient to lower energy significantly (Supplementary Information). In particular, the contact topology of final metastable states faithfully retains the network topology for $z_n\ageq 5.5$. 

To investigate the relation between the metastable energy landscape and the jamming transition, MS of distinct energy levels at $\phi=1$ are quasi-statically decompressed by shrinking all particles proportionally until the system approaches the jamming transition point \cite{Xu05,Desmond09}. 
We use an initial decompression step size of $\Delta \phi=0.01$ until the MS energy becomes zero, indicating that the configuration becomes unjammed. Iterative reduction of the step size then accurately finds the $\phi_c$ of the critical jamming state, defined as a configuration with $\epsilon_r$ between $10^{-16}$ and $2\times10^{-16}$, for which we also verify that the isostatic condition is fulfilled and that any further decompression leads to loss of all contacts. In these critical states, the soft particles can be exchanged for hard disks without changing the structure.

A main result of the present work is that the
critical packing fraction obtained in this way shows a strong negative correlation with the energy of the overjammed metastable state it was generated from. Thus, a wide range of $\phi_c$ can be efficiently constructed (Fig.~\ref{fig3}a,b). High energy MS at $\phi=1$ undergo the jamming transition at $\phi_c\approx0.84$, agreeing with common values of $\phi_{rcp}$ \cite{Bolton90, OHern03}. As MS energy decreases, the resulting critical packing fraction continuously increases, and the lowest-energy MS transition at $\phi_c\ageq 0.89$, significantly larger than typical values investigated before \cite{voivret2007space,Desmond09,du2023rearrangements}. The correlation between MS energy and $\phi_c$ does not change for different system size from $N=144$ to $N=1024$ (Fig.~\ref{fig3}b), while the maximum $\phi_c$ continues to increase with $N$. This suggests that at least as wide a range of $\phi_c$ can be realized in the limit of large system size. This result is consistent with several previous studies that the jamming transition occurs at a continuous range of packing fractions and it strongly depends on the preparation protocol both in 2D and 3D \cite{Chaudhuri10,Bertrand16,Ozawa17}. The generalized swap algorithm presented here, however, provides a far more efficient way of constructing critically jammed configurations of exceptionally high packing fraction. 

It is important to emphasize that the critically jammed states represent an arrangement of particles with unchanged given sizes $\{A_i\}$, even though the algorithm transiently replaces them with altered disk areas $\{A^{\prime}_i\}$. Had we constructed jammed states from the $\{A^{\prime}_i\}$, even higher area fractions could have been obtained (see Supplementary Information), due to the local size optimization. By contrast, we here present unusually dense packing solutions for the practical problem of jamming a given set of particles. 

What sets high-$\phi_c$ polydisperse packings apart structurally? Many traditional measures of amorphous disorder fail to distinguish them from lower-$\phi_c$ structures. 
Using the numbers $n_j$ of nearest neighbors of particle $j$ in the radical tessellation of the packing, 
the hexagonal orientation order parameter \cite{chaikin1995principles} is
   $ \Psi_{6,j}=\frac{1}{n_j}\sum_{k}\exp{6i\theta_{jk}},$
where $\theta_{jk}$ is the angle of the line connecting the positions of disks $j,k$ with respect to a reference direction. 
The average of $\left|\Psi_{6,j}\right|$ over all particles, shown 
in Fig.~\ref{fig3}c, increases very weakly with $\phi_c$. 
Likewise, the coefficient of variation of the tessellation topology, $c_n=\sqrt{\frac{1}{N}\sum_{j=1}^{N}(n_j-6)^2}/6$, a successful indicator of disorder in other contexts \cite{classen2005hexagonal,Kim18}, is insensitive to $\phi_c$ (Fig.~\ref{fig3}d). The radial distribution, $g(r)$, also shows no discernible change across jamming configurations with distinct $\phi_c$, making it an unsuitable measure for distinguishing them (Supplementary Information). 

\begin{figure}[!ht]
\centering
\includegraphics[width=.47\textwidth]{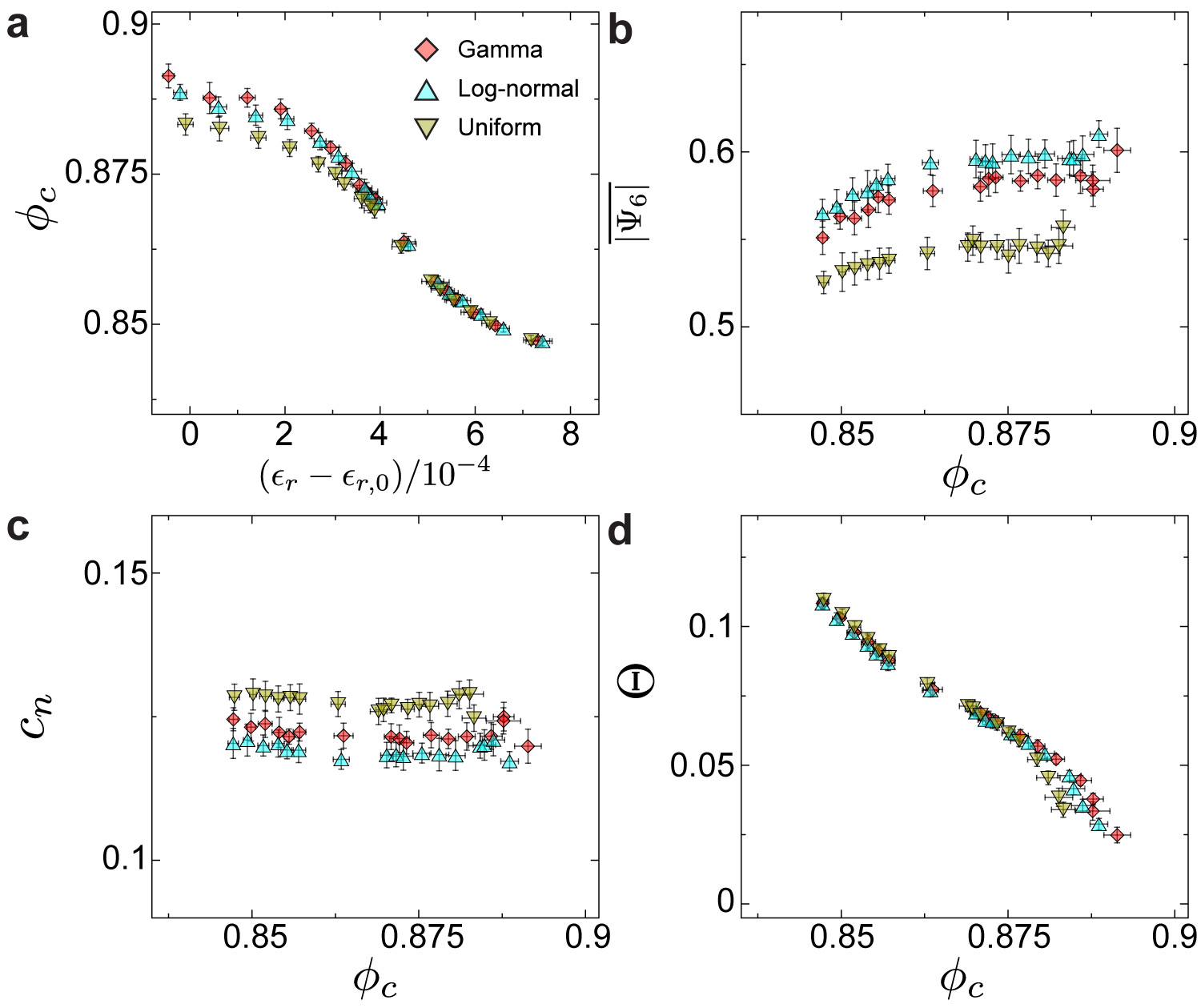}
\hfill
\caption{\textbf{a} Relation between metastable state energy of overjammed packings and critical packing fraction for different area distribution shapes of the polydisperse systems. Only slight quantitative changes result, confirming the conclusions of Fig.~\ref{fig3} for dependence on \textbf{b} hexagonal orientational order parameter, \textbf{c} coefficient of variation of the topology distribution, and \textbf{d} angle order parameter.}
\label{fig4}
\end{figure} 

By contrast, quantifying the local close-packing of disks directly provides a meaningful structural indicator. The angle order parameter $\Theta$ \cite{Tong18,Tanaka19,Kim22} averages the absolute differences $|\theta_{jkl}- \theta_{jkl,0}|$ for all triplets of particles that are mutual neighbors by radical tessellation, where $\theta_{jkl,0}$ is the angle of locally sterically optimal packing, with all three particles in contact, and $\theta_{jkl}$ is the actual angle within jammed particle packings, i.e., 
\begin{equation}
\label{eqtheta}
\Theta = \frac{1}{N}\sum_{j=1}^{N} \frac{1}{n_j}\sum_{k,l} \left|\theta_{jkl} - \theta_{jkl,0}\right|\ \,,
\end{equation}
see Supplementary Information for details. Figure~\ref{fig3}e confirms that this measure, which was shown to be an excellent predictor of MS energy at $\phi=1$ \cite{Kim22}, also correlates strongly with $\phi_c$ of critical jamming states.

All findings are robust with respect to changing the size distribution shape: Figure~\ref{fig4} compares results for Gamma and Log-normal distributions (two unimodal distributions), and by contrast for a uniform area distribution, all at $c_A=0.4$. While the maximum $\phi_c$ for the uniform distribution is slightly lower, and the quantitative measures of disorder slightly different, all trends are the same and in particular the angle order parameter $\Theta$ predicts $\phi_c$ universally.
Similar conclusions hold when $c_A$ is varied in the range of interest (see Supplementary Information for details), demonstrating again that this approach produces very dense packings for a wide variety of given size distributions. This is of interest as recent work shows that the phase behavior of glassy disk systems does depend on distribution shapes \cite{sampedro2020effect}.

\begin{figure}[!t]
\centering
\includegraphics[width=.47\textwidth]{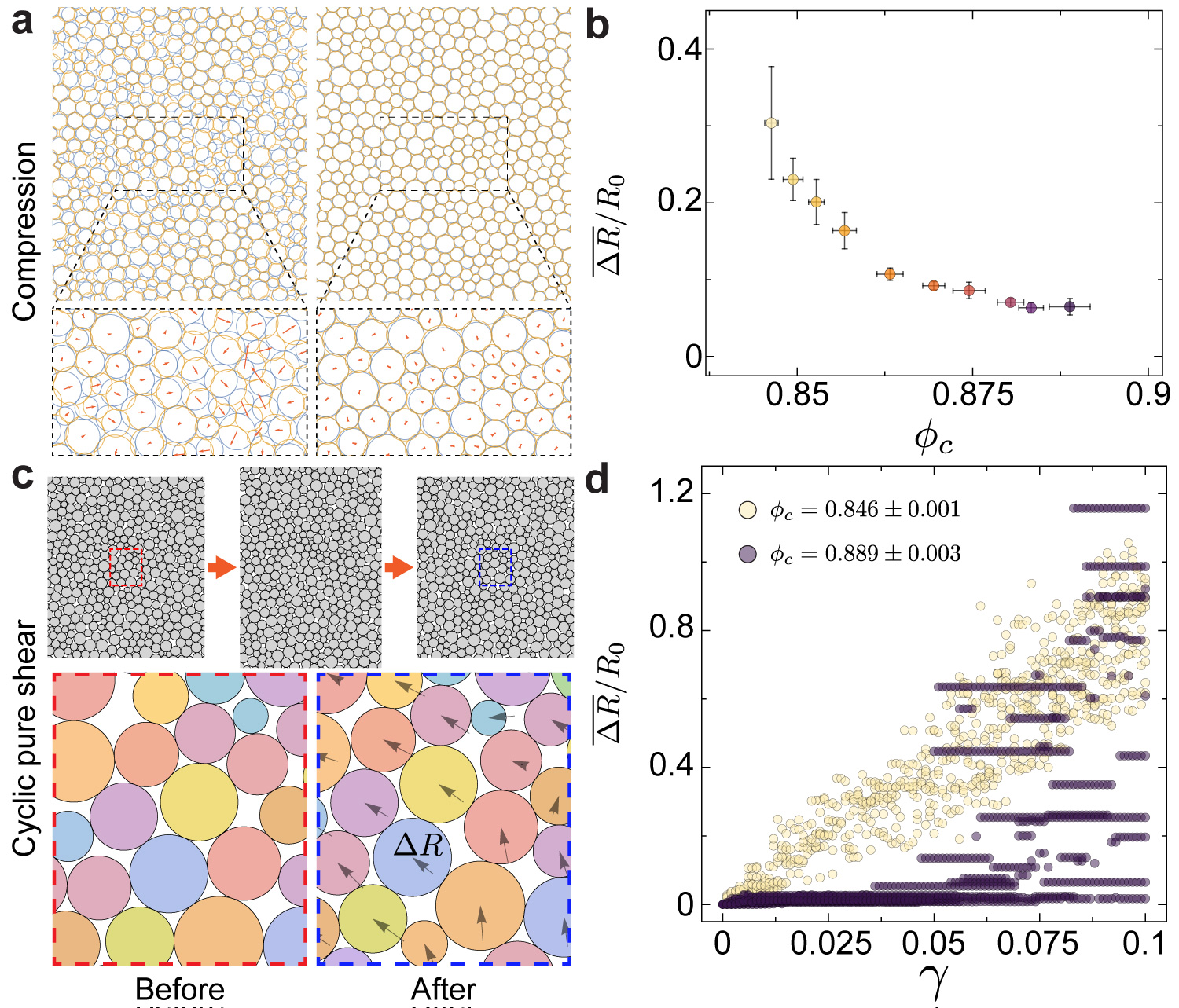}
\hfill
\caption{\textbf{a} Examples of compression of a packing for a low-$\phi_c$ critical state (left) and a high-$\phi_c$ critical states (right), indicating displacements of particles (arrows). \textbf{b} Averaged normalized particle displacement during compression from $\phi_c$ to $\phi=1$. For configurations with the highest $\phi_c$, hardly any displacement occurs. 
\textbf{c} Schematic of a cyclic pure shear deformation at $\phi_c$, indicating net displacement of individual particles $\Delta R$ after a shear cycle. \textbf{d} Mean particle displacement as a function of applied strain $\gamma$ for low-$\phi_c$ jammed states (yellow) and high-$\phi_c$ jammed states (purple). }
\label{fig5}
\end{figure}

We find that the structures of high-$\phi_c$ critical states are remarkably resilient to material-scale deformation. 
To assess how materials respond to volumetric changes, jamming configurations at $\phi_c$ are re-compressed up to $\phi=1$, and particle center displacements $\Delta R$ are monitored accordingly. Since the compression protocol involves proportionally increasing particle sizes rather than reducing the periodic box size, the $\Delta R$ represent non-affine motion. The average over all particles $\overline{\Delta R}/R_0$  ($R_0$ is the radius of a mean-area particle at $\phi=1$, and all rattlers are excluded) is a measure of structural rearrangement.  
We observe that structures unjamming at the highest $\phi_c$ have $\overline{\Delta R}/R_0 \ll 1$, while those unjamming at the lowest $\phi_c$ show substantial rearrangement (Fig.~\ref{fig5}a,b). Thus, the particles in extraordinarily low-energy overjammed states move almost perfectly affinely under uniform compression, maintaining their specific structure as the packing fraction changes. Very similar results are observed during the decompression process from $\phi=1$ to $\phi_c$ (Supplementary Information), suggesting that structures with high $\phi_c$ reliably maintain their underlying structure during volumetric changes. 

As a complementary test of structural robustness, 
a cyclic shear deformation is applied to each critically jammed configuration and its evolution from an initial configuration is evaluated (Fig.~\ref{fig5}c). Pure shear strain is first applied quasi-statically with an increment of $0.001$ until a target strain $\gamma\leq 0.1$ is reached. Then the system is quasi-statically sheared back to a state of zero strain. If a configuration unjams during shear, it is excluded from the analysis.
For each target strain $\gamma$, the mean particle displacement $\overline{\Delta R}/R_0$ is determined. Figure~\ref{fig5}d illustrates that low-$\phi_c$ structures readily rearrange under cyclic shear with displacements roughly proportional to $\gamma$; they are only marginally stable \cite{Arceri21}. Structures with extraordinarily high $\phi_c$, however, display qualitatively different behavior: they show negligible rearrangements until substantially large $\gamma$ values. Even after a finite displacement occurs, the system maintains this new configuration over a finite range of higher $\gamma$. The high-$\phi_c$ jammed states have thus inherited the structural resilience of the low-energy overjammed MS from which they were constructed -- those metastable states of low energy exhibit ultrastable characteristics, reminiscent of ultrastable glasses \cite{Swallen07,Guo12,Singh13,Ozawa18,Yeh20,Yanagishima21,Kim22}.

\begin{figure}[!t]
\centering
\includegraphics[width=.47\textwidth]{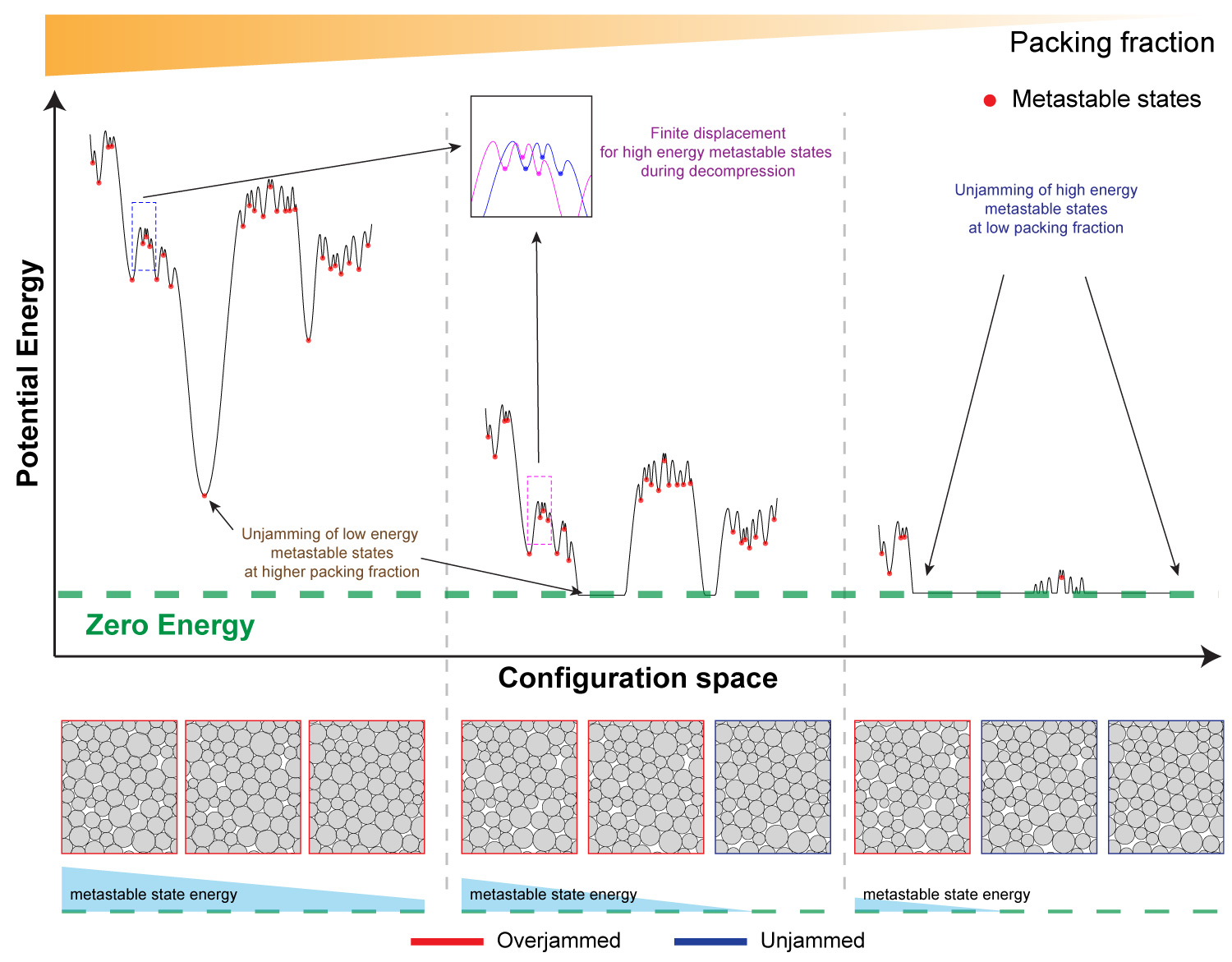}
\hfill
\caption{Illustration of qualitative changes in the energy landscape upon decompression of an overjammed soft-disk system ($\phi$ decreases left to right). Configurations representing the lowest-energy minima remain nearly unchanged, but unjam first.}
\label{fig6}
\end{figure}

Here we have demonstrated an efficient way to construct packings of polydisperse hard disks with unusually high density. We find that these extraordinary structures are also exceptionally robust against shear and compression -- indeed they inherit mechanical resilience from the overjammed MS they are constructed from because their relative particle positions are virtually unchanged from the overjammed states.
Our results suggest that reducing the packing fraction of overjammed packings during the decompression process maintains the shape of the energy landscape around its deepest minima, mainly shifting the landscape down along the energy axis (Fig.~\ref{fig6}), while higher minima do change their shapes and positions. As $\phi$ reaches the highest critical jamming fraction, the lowest minimum reaches zero energy and is lost. Further reduction of $\phi$ is necessary to unjam higher-lying minima, but their identity along the configuration axis is not maintained as faithfully as that of the lowest-energy states. 

The jamming transition and its underlying structural features are closely connected to the understanding of the glass transition and  glass dynamics, with low-lying minima on the energy landscape providing information about deep glassy states of low entropy. Very recent work on 2D glassy dynamics systems with polydispersity uses variants of simultaneous swap algorithms to efficiently equilibrate these systems to high packing fraction and high stability \cite{ghimenti2024irreversible}, a direct indication that the present results have relevance in glassy systems and that the rare, disordered high-$\phi_c$ configurations may represent analogs of minimal entropy, ideal-glass states \cite{royall2018race}. The robustness of our results against changes in the particle size distribution also hints at the universal character of these structures. Their further study, and the construction of analogous configurations in three dimensions, where studies of polydispersity effects are of great recent interest \cite{Hagh21,anzivino2023estimating},
will allow systematic exploration of disordered states by packing density. Extraordinary structures being correlated with extraordinary material properties, this has potentially far-reaching applications in the design of specific mechanical, electrical, or thermal behavior in glasses and other amorphous materials.


\end{document}


\maketitle

\section*{Definition of angle order parameter}

\begin{figure}[htb]
\centering
\includegraphics[width=.95\textwidth]{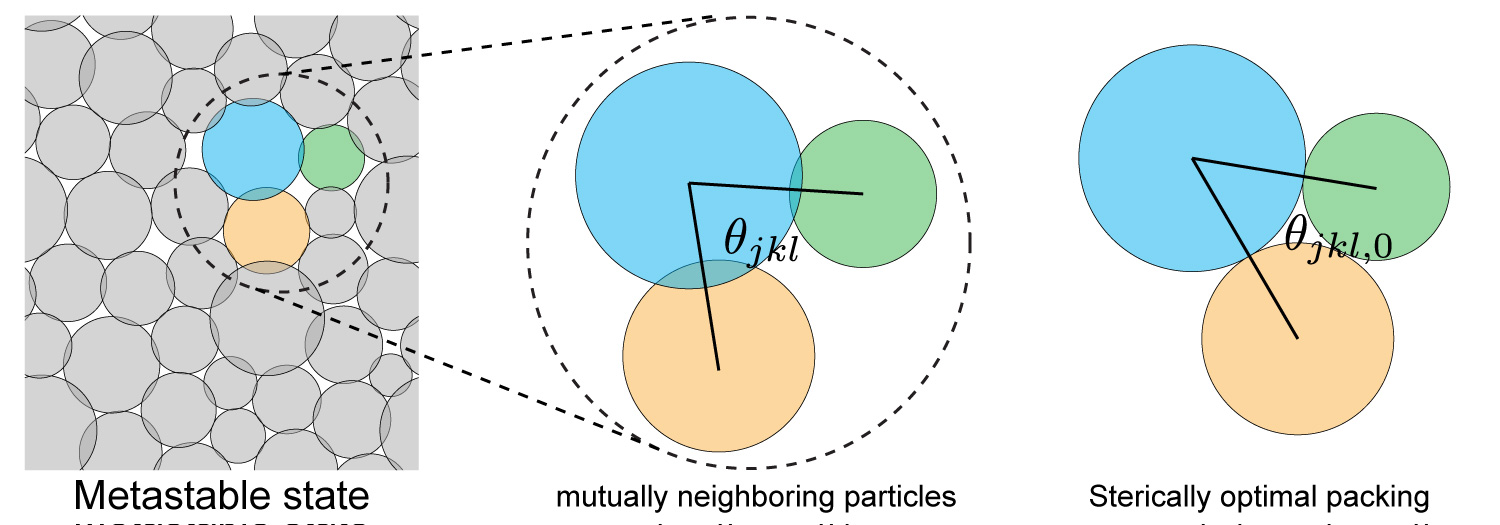}
\hfill
\caption{Schematics of a triplet of particles that are mutually neighbors within a metastable state (left and middle), and its corresponding sterically optimal packing. (right)}
\label{figS5}
\end{figure} 

To quantify structural closeness of particle packings to the reference state of the sterically optimal packing, we employ an angle order parameter. For each particle $j$ with $n_j$ neighbors determined by the radical tessellation, all pairs of neighboring particles, denoted as $k$ and $l$, are selected, where particles $k$ and $l$ are also neighbors. For every triplet of particles $j$, $k$, and $l$, the angles formed by a line connecting the centers of particles $j$ and $k$ and another line connecting the centers of particles $j$ and $l$, $\theta_{jkl}$, are measured (Fig.~\ref{figS5}). A configuration in which these particles mutually touch each other serves as a reference configuration for the sterically optimal packing. In this optimal packing, the angle between the three particles, $\theta_{jkl,0}$, follows directly from the particle sizes. The angle order parameter for each particle is computed as the average of the absolute difference between $\theta_{jkl}$ and $\theta_{jkl,0}$, namely $\Theta_j=\frac{1}{n_j}\sum_{k,l}\left|\theta_{jkl}-\theta_{jkl,0}\right|$. The angle order parameter for the particle packing is defined as the average of the angle order parameters of individual particles. 
\begin{equation}
    \Theta=\frac{1}{N}\sum_{j=1}^N\Theta_j=\frac{1}{N}\sum_{j=1}^N\frac{1}{n_j}\sum_{k,l}\left|\theta_{jkl}-\theta_{jkl,0}\right|\,.
\end{equation}
    
\section*{Impact of size distribution shapes on the effectiveness of the algorithm}

\begin{figure}[htb]
\centering
\includegraphics[width=.7\textwidth]{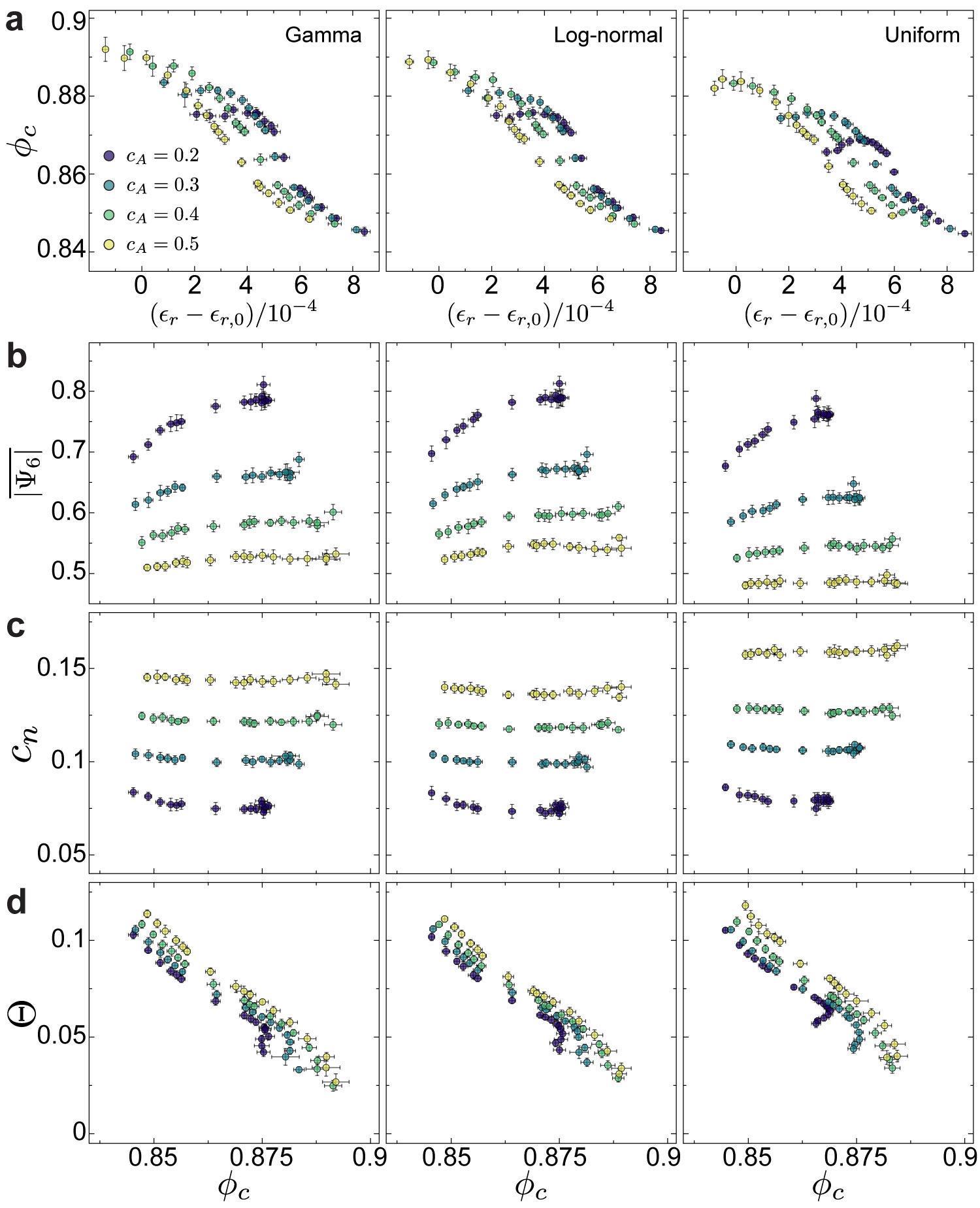}
\hfill
\caption{\textbf{a} A correlation between energy level of metastable states at $\phi=1$ and their corresponding critical packing fraction, $\phi_c$. Correlations between $\phi_c$ and structural measures, \textbf{b} Hexagonal orientation order parameter $\overline{\left|\Psi_6\right|}$, \textbf{c} coefficient of variation of radical tessellation topology $c_n$, and \textbf{d} the angle order parameter $\Theta$. In all plots, symbol color distinguishes polydispersity by coefficient of area variation $c_A$ (see \textbf{a})}
\label{figS1}
\end{figure} 

To assess the impact of size distribution shapes on the algorithm's efficiency, three different distributions - Gamma, Log-normal, and uniform distributions - are tested across a size polydispersity range of $0.2<c_A<0.5$. Regardless of the distribution shape and width, the algorithm robustly generates a broad range of critical packing fractions, $\phi_c$ (Fig.~\ref{figS1}a). While conventional structural measures such as $\overline{\left|\Psi_6\right|}$ and $c_n$ show weak correlations with $\phi_c$ (Fig.~\ref{figS1}b,c), the angle order parameter $\Theta$ is strongly predictive of $\phi_c$(Fig.~\ref{figS1}d), suggesting that closeness to sterically optimal packing significantly influences the jamming point. 

\section*{Particle displacement along the annealing process}

\begin{figure}[htb]
\centering
\includegraphics[width=.6\textwidth]{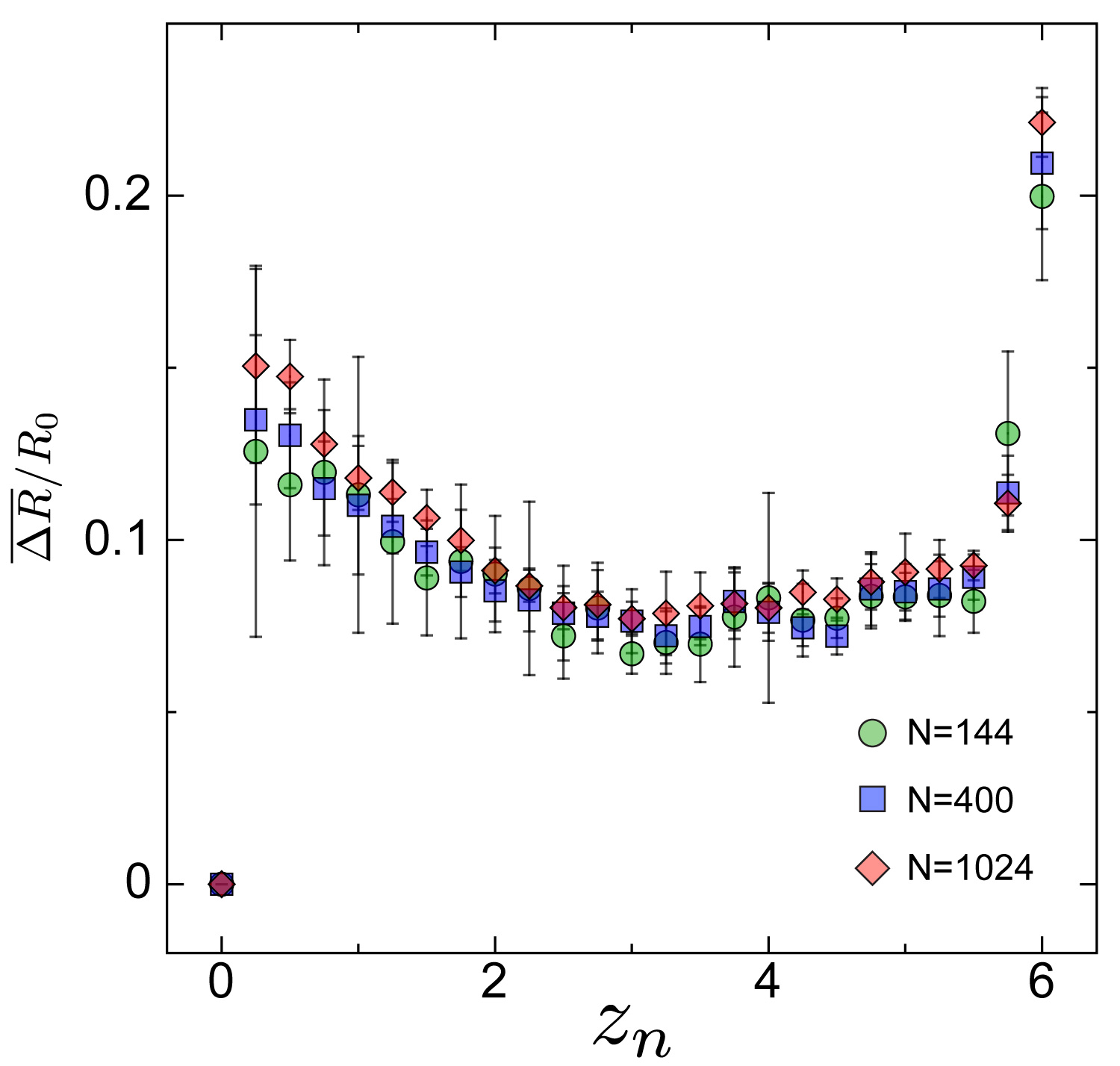}
\hfill
\caption{Mean normalized displacement of particles in overjammed packings along the annealing protocol by the simultaneous swap.}
\label{figS2}
\end{figure} 

The algorithm we propose here performs a single simultaneous size swap, which transitions high energy metastable states to new metastable states of adjustable energy level (the energy of these final metastable states being controlled by $z_n$). This annealing process involves adjustments of particle positions. To quantify the extent of particle motion, we compute the normalized displacement between an initial and a final metastable state following the simultaneous swap and energy minimization. Regardless of $z_n$, the average normalized displacement remains significantly less than 1 (less than a typical particle radius in absolute terms), indicating that local adjustments of particle positions are sufficient to reduce the energy of metastable states from the highest to any arbitrary values, even to the lowest energy level. 

\newpage

\section*{Radial distribution function of critical jamming states of various $\phi_c$}

\begin{figure}[htb]
\centering
\includegraphics[width=.98\textwidth]{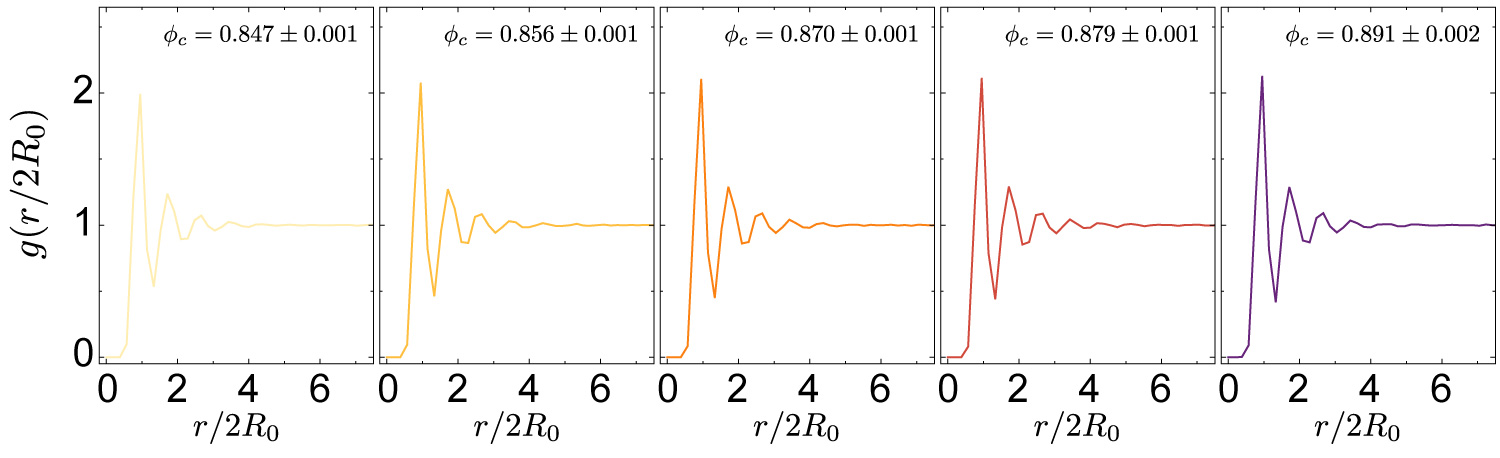}
\hfill
\caption{Radial distribution (pair correlation) function of critical jamming states at different $\phi_c$, showing no discernible differences. }
\label{figS3}
\end{figure} 

The radial distribution function, $g(r/2R_0)$ is computed to examine the distinct structural features of critical jamming states across various $\phi_c$. While the first peak value of $g(r/2R_0)$ slightly increases with increasing $\phi_c$, overall, $g(r/2R_0)$ exhibits minimal variation from the lowest $\phi_c$ to the highest $\phi_c$ (Fig.~\ref{figS3}). Therefore, we conclude that the radial pair correlation function of particle positions is unsuitable for distinguishing critical jamming states across various $\phi_c$. 

\newpage

\section*{Efficiency of the annealing algorithm for distinct $\phi$}

\begin{figure}[htb]
\centering
\includegraphics[width=.98\textwidth]{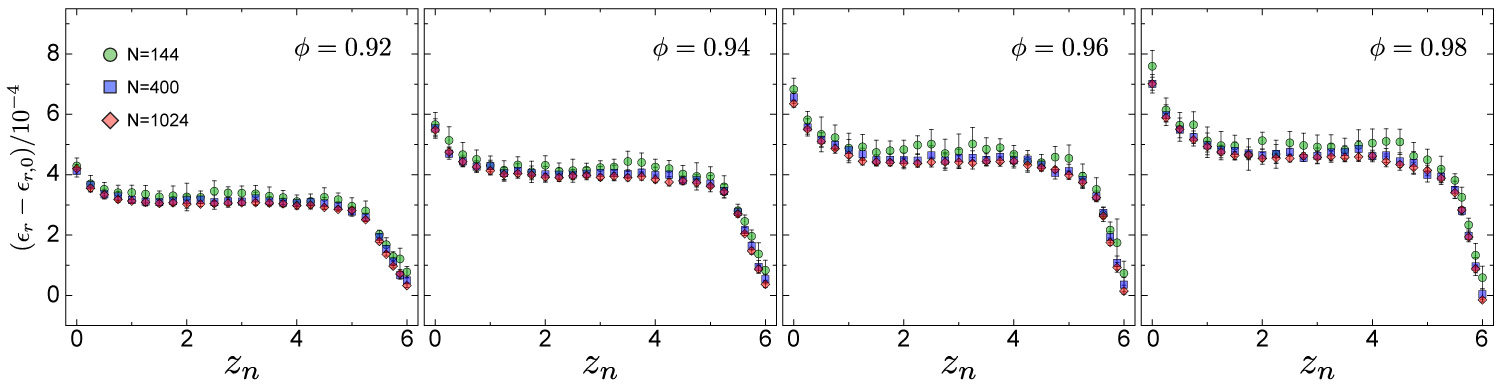}
\hfill
\caption{A correlation between average connectivity of network, $z_n$, and metastable state energy $\epsilon_r$ for different volume fraction $\phi$.}
\label{figS4}
\end{figure} 

Although the algorithm is primarily applied to generate metastable states of distinct energy levels at $\phi=1$ in this work, it can also be utilized to access a broad range of metastable states energies for any arbitrary $\phi>\phi_c$. To evaluate the effectiveness of the algorithm at different $\phi_c$, it is tested for four different overjammed packing fractions, $\phi=0.92,0.94,0.96$, and $0.98$. For each packing fraction, the algorihtm consistently generates metastable states covering the entire range from the highest to the lowest energy level, with the network coordination number $z_n$ remaining a valid control parameter (Fig.~\ref{figS4}).

\newpage

\section*{Displacement during the decompression process}

\begin{figure}[htb]
\centering
\includegraphics[width=.8\textwidth]{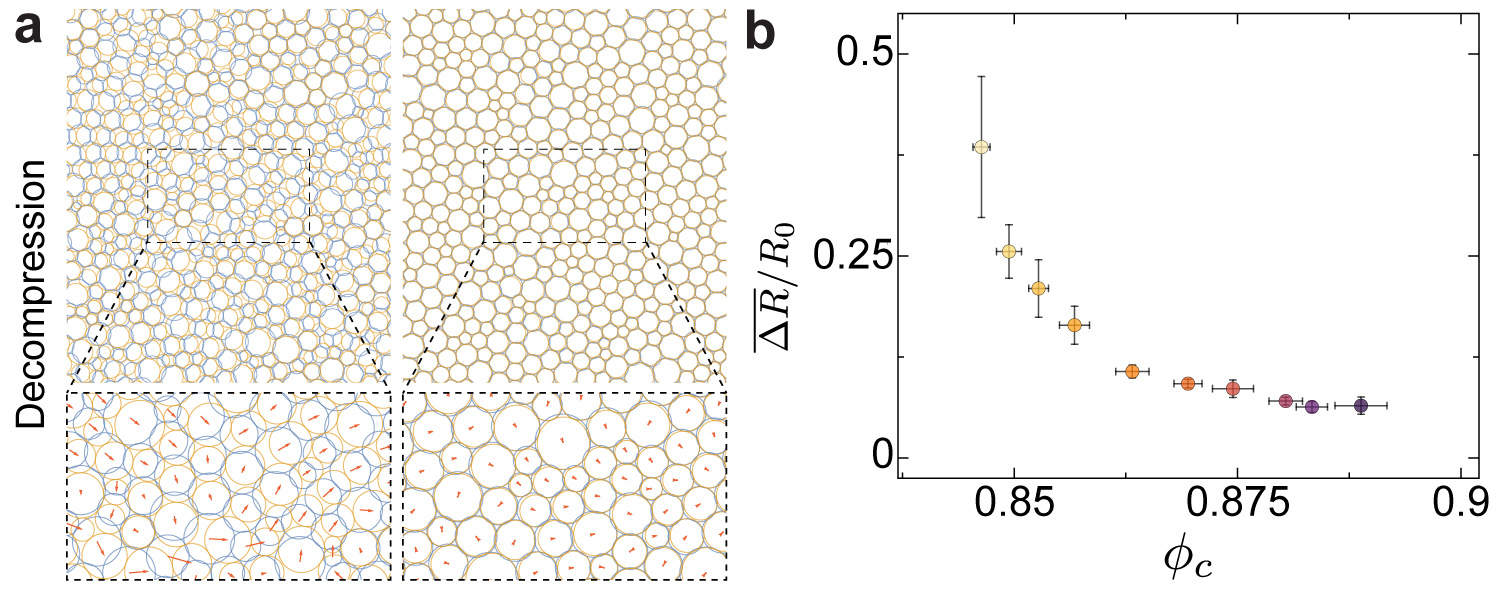}
\hfill
\caption{\textbf{a} Two examples of the decompression of packings for low (left) and high (right) $\phi_c$, with arrows indicating displacements along the decompression. \textbf{b} Mean normalized displacement of particles along the decompression process for different initial configurations of final $\phi_n$ after the decompression. }
\label{figS6}
\end{figure} 

As a supplementary displacement measure for the compression process outlined in the manuscript, we also calculate the normalized displacement during the decompression process from $\phi=1$ to $\phi_c$. Similar to the compression process, high energy metastable states (equivalently low-$\phi_c$ structures) exhibit high levels of displacement during the decompression process while low energy metastable states (equivalently high-$\phi_c$ structures) show small displacements, much smaller than a typical particle size. This further supports the conclusion that low-energy metastable states robustly maintain their structures during mechanical perturbations, whereas high energy metastable states do not. 

\newpage

\section*{Impact of adjusting the size distribution on critical packing fraction}

\begin{figure}[htb]
\centering
\includegraphics[width=.57\textwidth]{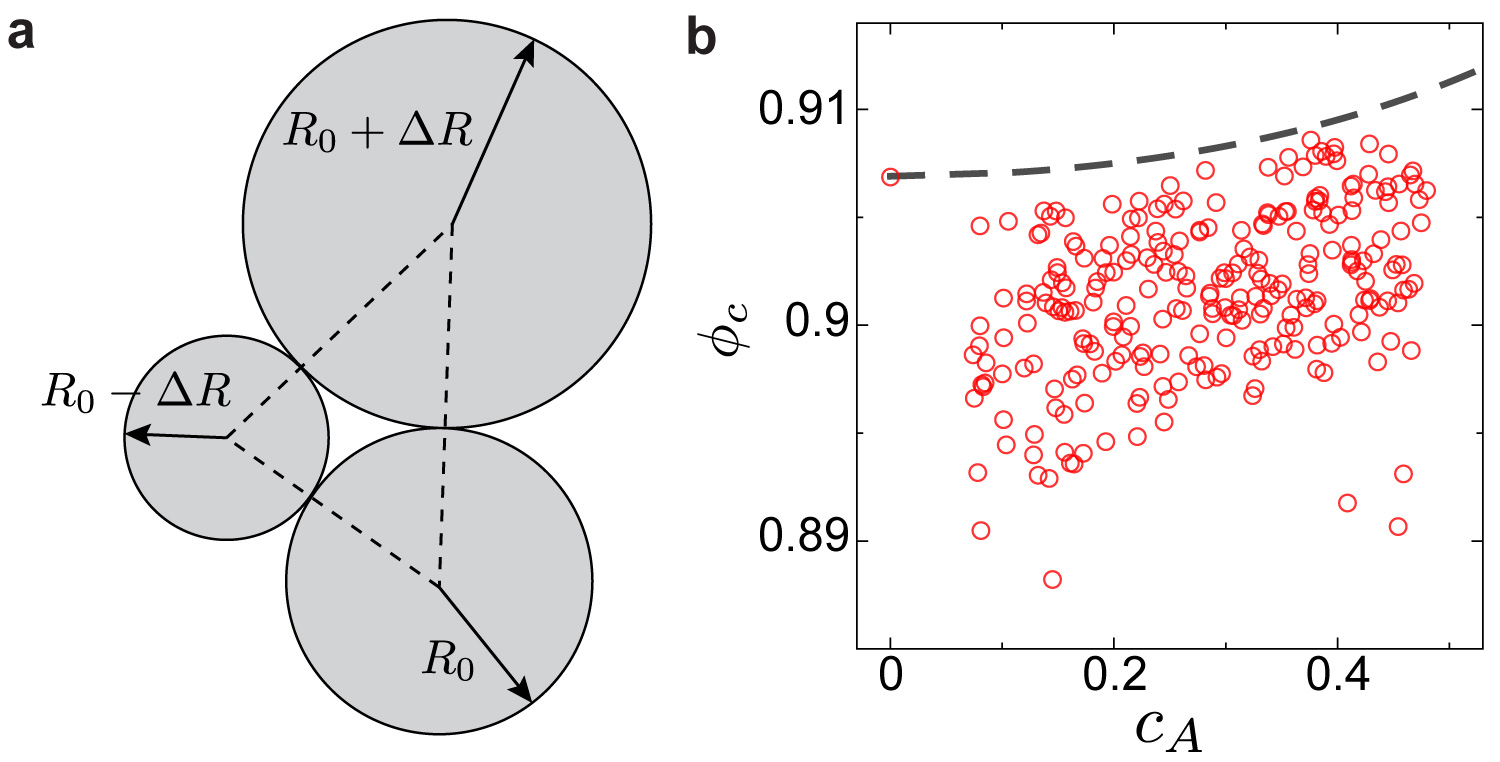}
\hfill
\caption{\textbf{a} Schematic of a representative locally optimal packing for three particles with a given size variation $\Delta R$. \textbf{b} Critical packing fractions, $\phi_c$, for structures retaining the modified set of particle sizes after network energy minimization (red circles, system size $N=400$). Many samples approach and exceed the monodisperse crystalline fraction $\phi_{hex}=0.9069\dots$ for the significantly large size polydispersity. The gray dashed line indicates the theoretical estimate of the maximum $\phi_c$ based on the representative configuration shown in \textbf{a}. }
\label{figS7}
\end{figure} 
While our primary focus in this work is on the densest packings of a given size distribution (i.e. a fixed set of particle sizes $\{A_i\}$), even higher critical packing fractions $\phi_c$ of polydisperse disk packings can be achieved if adjustments to the size distribution are permitted. To investigate this, we adapt the original algorithm to retain the adjusted set of sizes after the network energy minimization, $\{A^{\prime}_i\}$, instead of reassigning the fixed particle sizes, $\{A_i\}$, in the order of $\{A^{\prime}_i\}$. This modification results in a subtly different size distribution from the original one. When applying this algorithm for $z_n=6$, we observe a significant increase in the accessible range of $\phi_c$ compared to the original cases (Fig.~\ref{figS7}b). 

This result is understandable as the adjusted sizes $\{A^{\prime}_i\}$ specifically optimize local packing (for $z_n=6$, these are sizes equivalent to those constructed by the Circle Packing algorithm). To estimate a theoretical limit $\phi_{c,max}$ for a given $c_A$, we consider a sterically optimal triplet packing (Fig.~\ref{figS7}a). This packing consists of one particle with an average size $R_0$, another particle with size $R_{-}=R_0+\Delta R=R_0(1-\Delta r)$, and the other particle with size $R_{+}=R_0+\Delta R=R_0(1+\Delta r)$. Assuming a Gamma distribution of particle sizes, the normalized deviation $\Delta r$ can be expressed in terms of the coefficient of variation of particle area, $c_A$, as follows:
\begin{equation}
    \Delta r=\frac{\sqrt{1-f(c_A)^2}}{f(c_A)}
\end{equation}

\begin{equation}
    f(c_A)=\frac{c_A\Gamma\left(c_A^{-2}+1/2\right)}{\Gamma(c_A^{-2})}
\end{equation}
The packing fraction of this optimal packing is calculated as the ratio of the sum of disk sectors within the triangle formed by joining the particle centers to the size of the triangle (Fig.~\ref{figS7}a). Expanding the angles of the triangle for small $\Delta r$, we can derive an estimate of $\phi_{c,max}$, namely 
\begin{equation}
    \phi_{c,max}(c_A)=\frac{\pi+\left(\frac{2\pi}{3}-2\sqrt{3}\right)(\Delta r)^2}{2\sqrt{3(1-(\Delta r)^2)}}
\end{equation}
This upper-bound estimate is displayed in Fig.~\ref{figS7}b and is consistent with the data. We emphasize that such particle-size adjustments are not typically accessible in practical packing problems, where a given set of particles is to be jammed. All results in the main text find rearrangements of such given sets of particle sizes $\{A_i\}$, for various distribution shapes and widths.